\documentclass[letter,scriptaddress,twocolumn,prl,showpacs,showkeys]{revtex4-1}
		\usepackage{epsfig}

\begin{document}

\preprint{}

\title{Reconstructing the nature of the first cosmic sources from 
the anisotropic 21-cm signal}

\author{Anastasia Fialkov}
\email{anastasia.fialkov@gmail.com}
\affiliation{Departement de Physique, Ecole Normale Superieure,
CNRS, 24 rue Lhomond, 75005 Paris, France}
\author{Rennan Barkana}
\affiliation{Raymond and Beverly Sackler School of Physics and Astronomy,
Tel Aviv University, Tel Aviv 69978, Israel}
\author{Aviad Cohen}
\affiliation{Raymond and Beverly Sackler School of Physics and Astronomy,
Tel Aviv University, Tel Aviv 69978, Israel}

\date{\today}

\begin{abstract}
  The redshifted 21-cm background is expected to be a powerful probe
  of the early Universe, carrying both cosmological and astrophysical
  information from a wide range of redshifts. In particular, the power
  spectrum of fluctuations in the 21-cm brightness temperature is
  anisotropic due to the line-of-sight velocity gradient, which in
  principle allows for a simple extraction of this information in the
  limit of linear fluctuations. However, recent numerical studies
  suggest that the 21-cm signal is actually rather complex, and its
  analysis likely depends on detailed model fitting. We present the
  first realistic simulation of the anisotropic 21-cm power spectrum
  over a wide period of early cosmic history. We show that on
  observable scales, the anisotropy is large and thus measurable at
  most redshifts, and its form tracks the evolution of 21-cm
  fluctuations as they are produced early on by Lyman-$\alpha$
  radiation from stars, then switch to X-ray radiation from early heating sources, 
  and finally to ionizing radiation from stars. In
  particular, we predict a redshift window during cosmic heating (at
  $z \sim 15$), when the anisotropy is small, during which the shape
  of the 21-cm power spectrum on large scales is determined directly
  by the average radial distribution of the flux from X-ray sources.
  This makes possible a model-independent reconstruction of the X-ray
  spectrum of the earliest sources of cosmic heating.
  
\end{abstract}

\pacs{98.80.Es, 95.75.Pq, 98.58.Ge}
\keywords{21-cm signal}

\maketitle

{\it \bf Introduction.}\ \ The high-redshift intergalactic medium
(IGM) can be probed by measuring the intensity of photons in the
redshifted 21-cm line of the hydrogen atom
\cite{Barkana:2006,2001Review,LoebBook}. The 21-cm intensity
(expressed as a brightness temperature) contains a mixture of
astrophysical and cosmological information \cite{MMR} from the era
when the Universe was mostly neutral, i.e., before and during the
Epoch of Reionization (redshift $z \gtrsim 7$).  Because it is coupled
to numerous inhomogeneous inputs, the 21-cm signal fluctuates on all
scales. Among the most important sources of perturbations are those in
the matter density, $\delta$, the peculiar velocity, and radiative
backgrounds produced by stars and their remnants; the latter include
X-rays which heat the IGM, Ly-$\alpha$ photons which couple the 21-cm
line to the gas temperature, and ionizing photons. The various
fluctuation sources dominate at different epochs throughout cosmic
history, leading to a succession of peaks in the three-dimensional
power spectrum of the 21-cm brightness temperature fluctuations
(hereafter the "21-cm power spectrum") averaged over the line of sight
\cite{Barkana:2005b,Pritchard:2007,21cmfast}. However, the exact size
and scale-dependence of the 21-cm fluctuations are still highly
uncertain, since they depend on the properties of high-redshift
astrophysical sources, which are poorly constrained due to the lack of
observations at present.

Of particular interest is the nature of the first heating sources
which raised the gas temperature above the temperature of the cosmic
microwave background (CMB) radiation. The candidates for these sources
are varied and include X-ray binaries, mini-quasars, and hot gas in
the first galaxies, plus more exotic possibilities such as decaying
dark matter. The spectral energy distribution (SED) of X-ray photons
emitted in each case can be very different, ranging from a soft
power-law spectrum \cite{Furlanetto:2006} to a hard spectrum that
peaks at energies of several keV as in the case of X-ray binaries
\cite{Mirabel:2011,Fragos:2013}. Because the heating is inhomogeneous,
the character of the SED is imprinted in the 21-cm power spectrum,
which can be used to constrain the average energy of the X-ray photons
which heat the IGM \cite{Fialkov:2014,Pacucci:2014}.

Since 21-cm fluctuations arise from a superposition of several
different sources of fluctuations, the signal is complex and its
interpretation requires detailed modeling. Finding additional probes
of the effects of various astrophysical objects would be a major step
forward, especially more direct, model-independent ways to reveal
their properties. One possible tool is the directional dependence of
the 21-cm fluctuations.

{\it \bf Anisotropy of the power spectrum.}\ \ The 21-cm power
spectrum is predicted to be anisotropic due to the radial component of
the peculiar velocity gradient created by structure formation
\cite{Kaiser:1987,Bharadwaj:2004,Barkana:2005a}.  Specifically, in a
region with a lower velocity gradient than average, a larger stretch
of atoms along the line of sight contributes to the optical depth of
the 21-cm line. Within linear theory, this effect results in a power
spectrum that is a simple sum of terms \cite{Barkana:2005a},
\begin{equation}
  P({\bf k},z) = P_{\mu^0}(k,z)+3P_{\mu^2}(k,z)\mu_k^2+5P_{\mu^4}(k,z)\mu_k^4\ ,
\label{Eq:Pk}
\end{equation} 
where $\mu_{\bf k} \equiv \cos \theta_{\bf k}$ in terms of the angle
$\theta_{\bf k}$ between the wavevector ${\bf k}$ of a given Fourier
mode and the line of sight \cite{FN1}.

The ${\bf \mu_k}$-dependence of the 21-cm power spectrum can in
principle be used to differentiate between the cosmological and
astrophysical contributions to the fluctuations \cite{Barkana:2005a}.
In particular, since the velocity gradient is directly related to
density through the continuity equation, $P_{\mu^4}$ is proportional
to the primordial density power spectrum $P_\delta$, $P_{\mu^0}$ is
the sum of all the isotropic 21-cm fluctuation sources (i.e., those
other than the velocity gradient), and $P_{\mu^2}$ is proportional to
the correlation of density with the isotropic 21-cm fluctuation
sources.  However, 21-cm fluctuations are non-linear, and recent
numerical investigations during cosmic reionization
\cite{McQuinn,Jensen,Shapiro:2013,MaoY} suggest that it will be
difficult to make this separation of terms and recover cosmological
information from the 21-cm power spectrum. 

These investigations focused on the $P_{\mu^4}$ term, which carries
the primordial cosmological information. This term is typically the
smallest, and is most easily contaminated by non-linearity
\cite{FN2}. In
what follows, we rekindle the importance and usefulness of the
linear-theory polynomial decomposition of the power spectrum
(eq.~\ref{Eq:Pk}) by focusing on the $P_{\mu^2}$ term instead.

{\it \bf Realistic mock data.}\ \ While upcoming 21-cm observations
will be hard-pressed to beat other probes in terms of basic cosmology,
they are likely to reveal to us several chapters in the history of
stars and galaxies for the first time. We use simulated 21-cm
observations for the first realistic study \cite{FN3}
of the anisotropy in the 21-cm power spectrum during the entire period
of cosmic history relevant for upcoming experiments, from the
formation of the first substantial population of stars (at $z\sim 30$)
until the end of reionization ($z\sim 7$). We create mock data using a
hybrid, semi-numerical simulation in which the non-linear evolution of
the redshifted 21-cm signal is followed accounting for its dependence
on $\delta$, the line of sight velocity gradient, and inhomogeneous
X-ray, Ly-$\alpha$ and ionizing radiative backgrounds (e.g.,
\cite{Fialkov:2014} and references therein). The simulated volume is
$384^3$ Mpc$^3$ with a 3~Mpc resolution, and assumes that stars form
via atomic cooling. In addition, we average our results over twenty
randomly generated sets of initial conditions to decrease the
statistical errors on large scales. On the resolved scales,
perturbations in the density and velocity gradient fields are linear,
but those in the galaxy density (and thus in the radiation and 21-cm
fields) are enhanced (``biased'') and thus non-linear \cite{Unusual}.
 
{\it \bf Results and discussion.}\ \ Among the various terms in
eq.~(\ref{Eq:Pk}), $P_{\mu^0}$ dominates the usually discussed total
(angle-averaged) 21-cm power spectrum, so we focus on $P_{\mu^2}$ as
the most promising additional source of observable information. We
find that $P_{\mu^4}$ is more difficult both to measure and to
interpret (so we mostly leave it for future work). Now, as noted
above, in linear theory, $P_{\mu^2}^{\rm lin}$ would be proportional
to the cross-correlation of density with all the isotropic 21-cm
fluctuation sources \cite{FN4}. While the total 21-cm power spectrum does track the
changing sources of fluctuations indirectly through changes of
magnitude and slope, $P_{\mu^2}^{\rm lin}$ does this much more
unambiguously and explicitly through changes of {\it sign}\/
\cite{Rich}. This is due to the fact that some sources are positively
correlated with density, and others are negatively correlated.

\begin{figure*}
\includegraphics[width=3.0in]{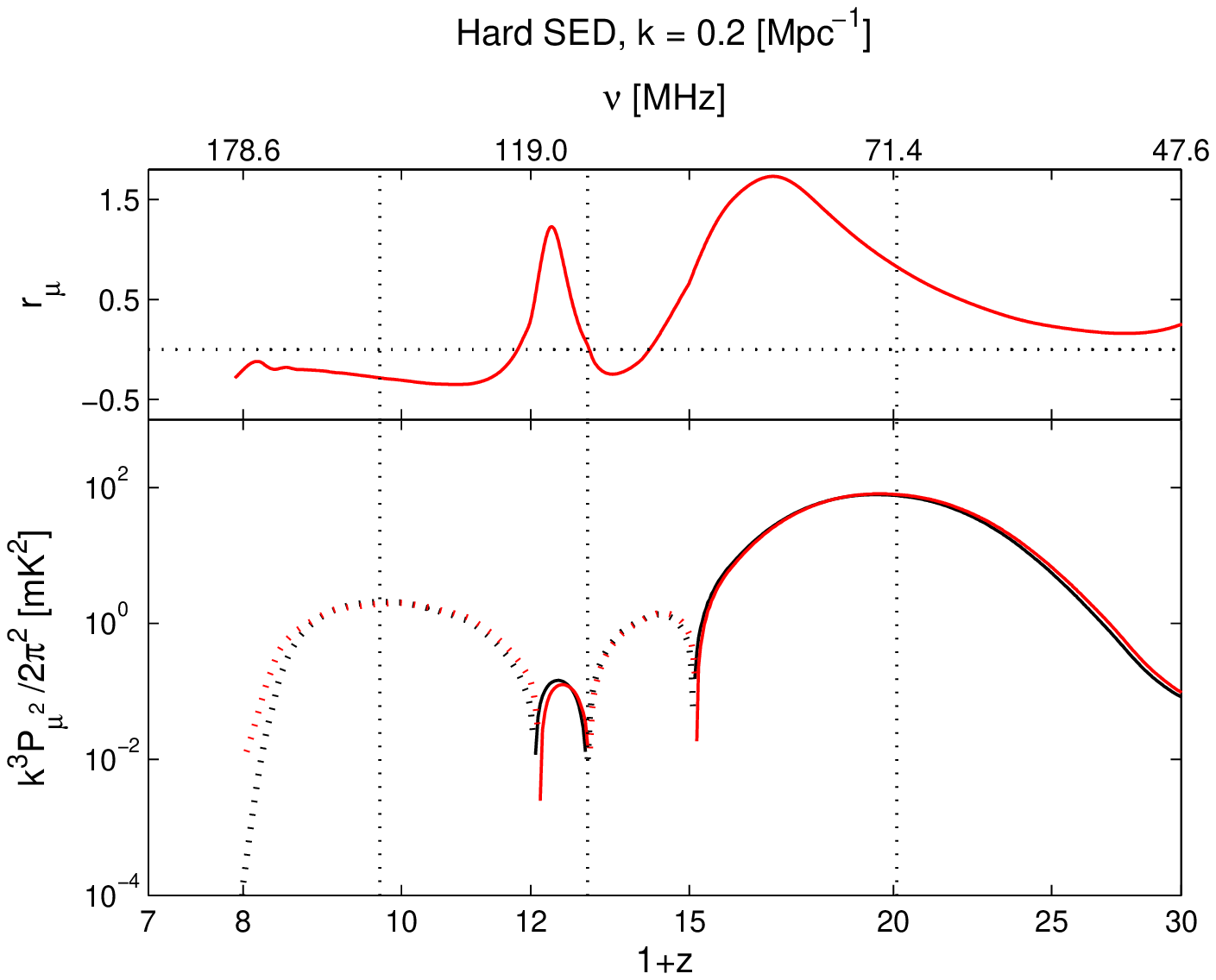}\includegraphics[width=3.0in]{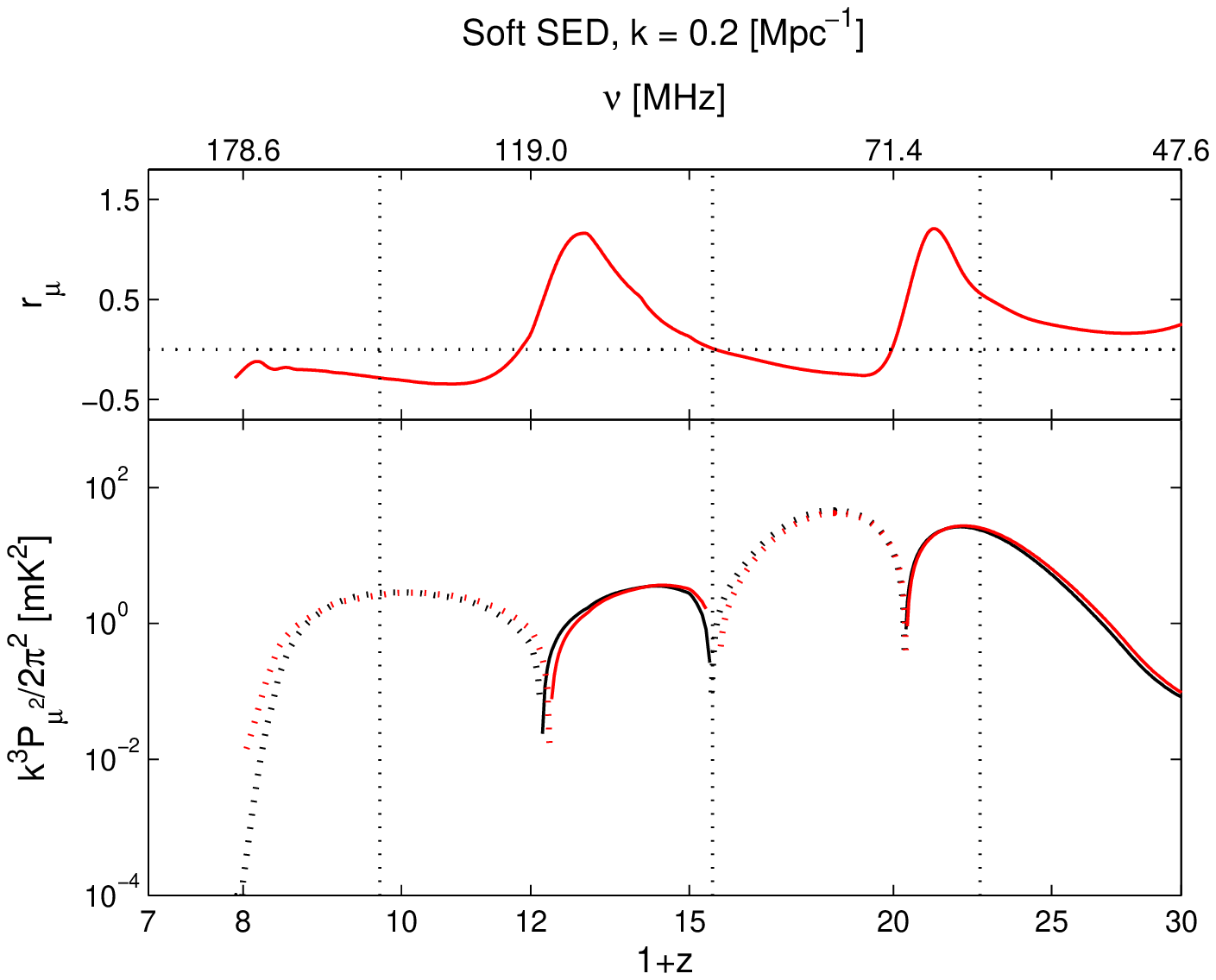}
\includegraphics[width=3.0in]{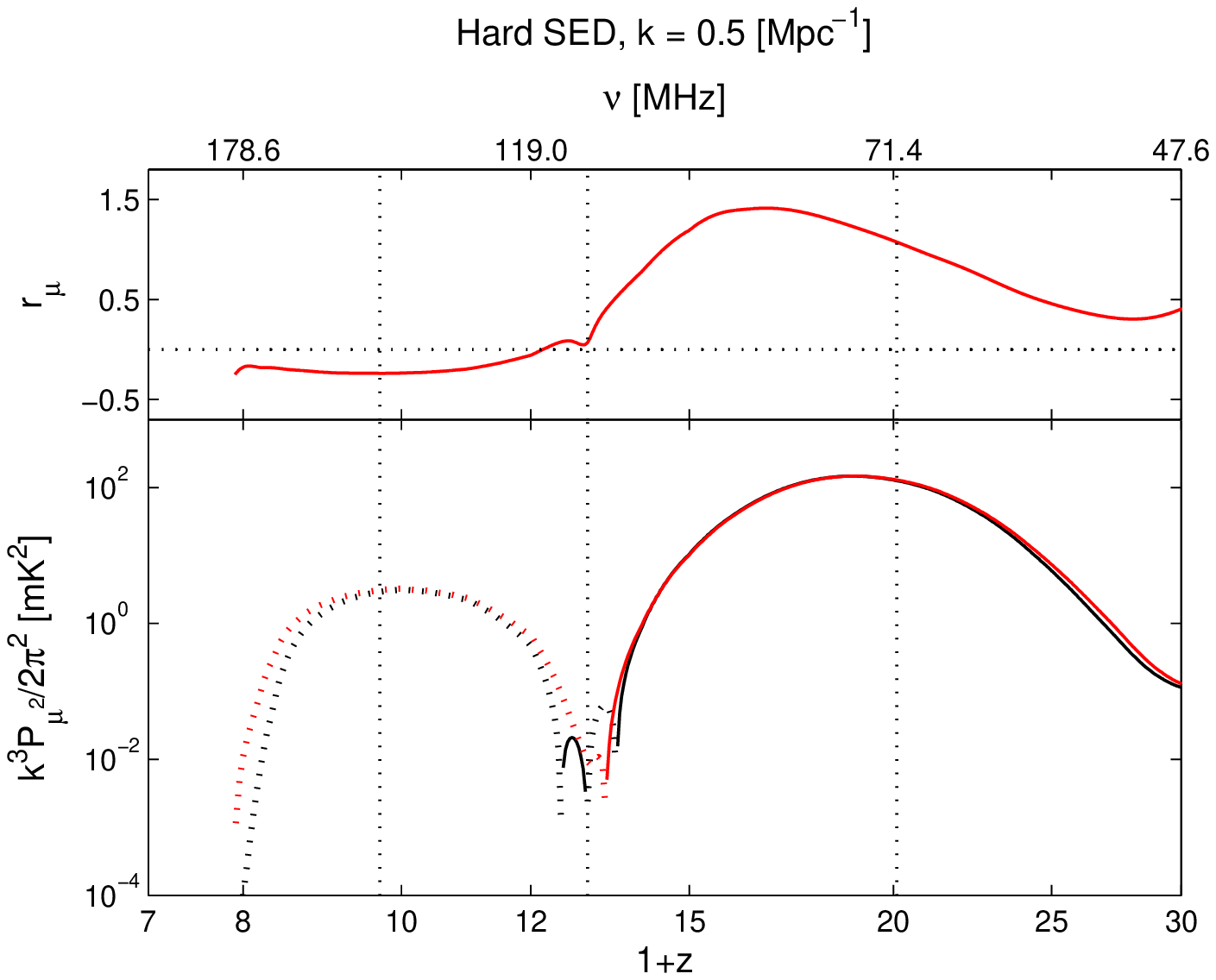}\includegraphics[width=3.0in]{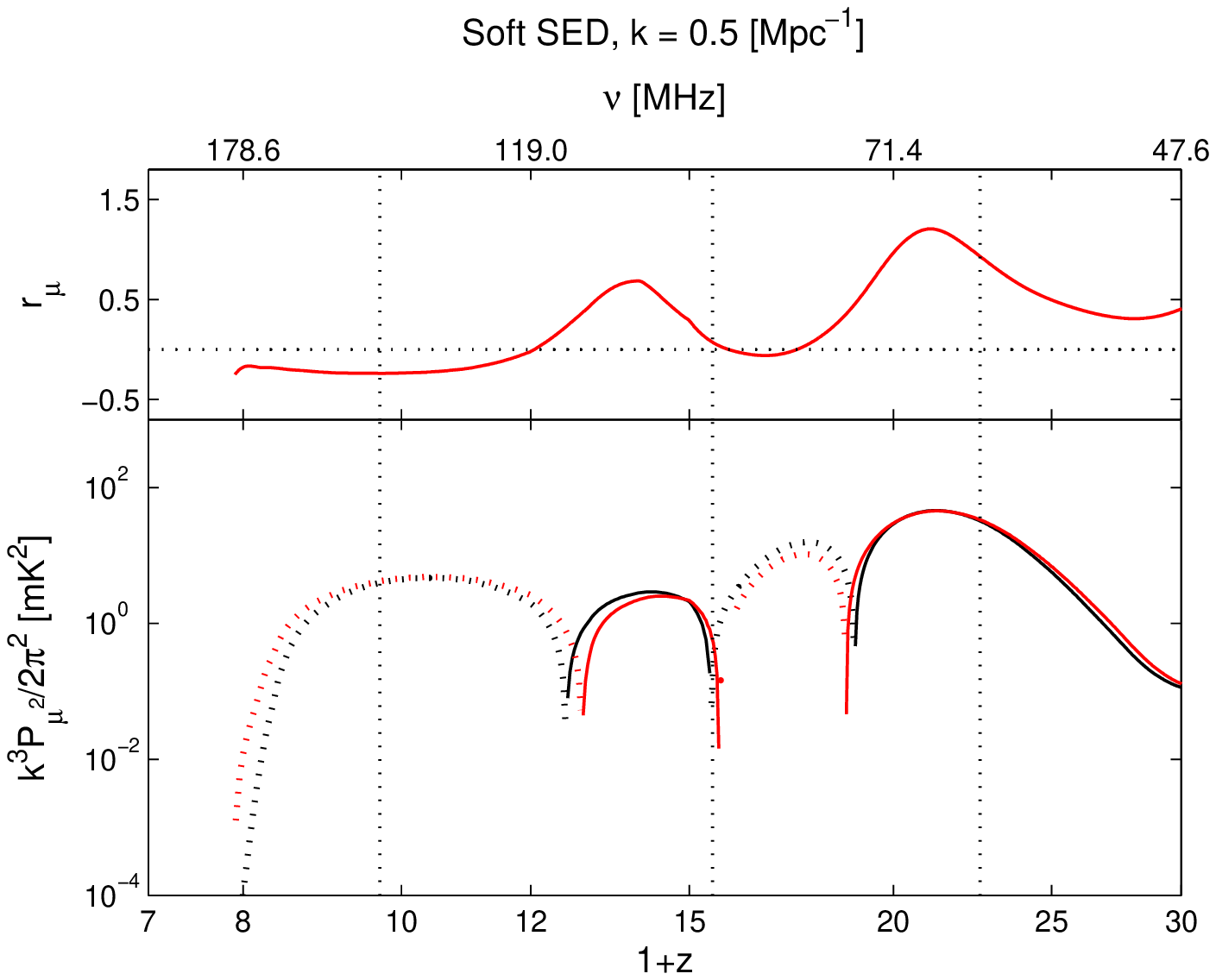}
\caption{The anisotropy ratio $r_\mu$ (top plot within each panel) and
  coefficient $P_{\mu^2}(k,z)$ (bottom plot within each panel).  For
  the latter, we compare the actual observable $P_{\mu^2}$ (red) to
  the theoretical result $P_{\mu^2}^{\rm lin}$ based on linear
  separation (black). The observed $P_{\mu^2}$ is shown only where the
  fit (to eq.~\ref{Eq:Pk}) that produced it works well (i.e.,
  $R^2>0.9$).  Positive values of $P_{\mu^2}$ or $P_{\mu^2}^{\rm lin}$
  are shown with solid lines, and negative with dotted lines. The
  results are shown for the cases of a hard X-ray SED (left panels) or
  a soft SED (right panels) at two wavenumbers, $k = 0.2$~Mpc$^{-1}$
  (top panels) and $k = 0.5$~Mpc$^{-1}$ (bottom panels). A thin
  horizontal line in each panel indicates $r_\mu=0$. The thin vertical
  lines in each panel mark global milestones in the history of the
  Universe; from left to right: the mid-point of cosmic reionization
  (i.e., when half the cosmic gas has been reionized), the redshift of
  the heating transition, and the peak of the Ly-$\alpha$ coupling era
  (specifically, the redshift of the peak fluctuation level at $k=0.3$
  during this era).}
\label{Fig:P}
\end{figure*}

$\bullet$ {\it Tracking cosmic history.}\ \ As shown in
Figure~\ref{Fig:P}, the actual observed $P_{\mu^2}$ (obtained from
fitting the form of eq.~(\ref{Eq:Pk}) to mock data) mostly tracks
$P_{\mu^2}^{\rm lin}$, as (going from early to late times) it is
positive during the epoch of Lyman-$\alpha$ fluctuations, goes
negative when X-ray heating fluctuations take over, then positive
again as heating fluctuations continue but the mean IGM temperature
rises above that of the CMB (a key cosmic milestone termed the
``heating transition''), and finally back to negative once ionization
fluctuations come to dominate \cite{FN5}.

We also consider a simpler, more robust measure of anisotropy that
does not require fitting the power spectrum to a particular form. We
define the "anisotropy ratio"
\begin{equation}
r_\mu(k,z)\equiv \frac{ \langle P({\bf k},z)_{|\mu_k|>0.5} \rangle}
{\langle P({\bf k},z)_{|\mu_k|<0.5}\rangle} -1\ ,
\label{Eq:rmu}
\end{equation} 
which indicates how much $P({\bf k},z)$ increases or decreases with
$\mu_k$, on average \cite{FN6}.  In particular, when the power spectrum has
a weak angular dependence, $r_\mu$ is close to zero; on the other
hand, when the power spectrum as a function of $\mu_k$ has a
considerable positive or negative tilt, the anisotropy ratio is far
from zero and its sign is positive or negative, respectively.

The first conclusion from considering the anisotropy ratio (which is
also shown in Figure~\ref{Fig:P}) is that the anisotropy is generally
large (i.e., $|r_\mu|$ is of order unity) and thus potentially
observable. (We additionally find in our simulation that its measurement is not very sensitive to the lowest
values of $|\mu_k|$, which are highly foreground contaminated
\cite{Morales:2012,Trott:2012,Liu:2014,Pober:2013}). More importantly,
$r_\mu$ is rich in information. Thanks to the characteristic structure
of peaks and plateaus, the timing of cosmic events can approximately
be read off the anisotropy ratio.

Indeed, the sign and general shape of $r_\mu$ mostly track those of
$P_{\mu^2}$, since the $P_{\mu^2}$ term usually dominates the
angle-dependent terms. There are interesting differences, though.
When $P_{\mu^2}$ passes through zero (as it is switching sign), this
usually indicates that two different fluctuation sources (one
positively and one negatively correlated with density fluctuations)
are superposing and canceling out (at a particular wavenumber), so
$P_{\mu^0}$ is also at that moment small compared to the velocity
gradient term, and $r_\mu$ has a significantly positive value
($P_{\mu^4}^{\rm lin}$ is always positive, and the non-linear $r_\mu$
reflects this). However, when $P_{\mu^2}=0$ due to the heating
transition, $r_\mu \sim 0$ as well since the heating fluctuations
continue to dominate through this transition and are not canceled out
by a second source. Note that the heating transition can be recognized
(at least at $k=0.2$~Mpc$^{-1}$, for which the heating fluctuations
clearly dominate around that time) as the point when $r_\mu=0$ as it
rises with time from negative to positive values.

While current observations point to a most likely scenario of heating
by X-ray binaries with a hard spectrum
\cite{Fragos:2013,Fialkov:2014}, there remains a great uncertainty
about X-ray sources at such high redshifts. To illustrate how the
spectrum of the X-ray sources can be probed, we also show results for
a soft power-law SED \cite{Furlanetto:2006} normalized to the same
total emitted energy. We focus on relatively large (i.e.,
cosmological) scales, and show two different wavenumbers in order to
illustrate the scale dependence.  The observable quantities
($P_{\mu^2}$ and $r_\mu$) reflect the fact that for the hard SED
(compared to the soft one), the heating transition occurs later, and
the heating fluctuations are smaller (especially on small scales), so
they dominate over other sources for a shorter time. The heating is
late since only a fraction of the energy of the hard X-rays is
absorbed, and the heating fluctuations are small since the hard X-rays
mostly come from large distances \cite{Fialkov:2014}.

$\bullet$ {\it Measuring the spectrum of early X-ray sources.}\ \
Recognizing our ability to track cosmic history using $P_{\mu^2}$
and/or $r_\mu$ makes possible an especially promising direct
measurement. Among the three cosmic periods considered above, the
X-ray heating era is the best target, since its source properties are
highly uncertain (unlike the Lyman-$\alpha$ spectrum, which is
predicted to be relatively similar for modern and primordial stars
\cite{Barkana:2005b}), and it is also relatively easy to interpret
(unlike the bubble structure of reionization that is a result of the
small mean-free-path of ionizing photons in the neutral IGM). Also,
during the era when heating fluctuations dominated the 21-cm
fluctuations there was a unique moment (the heating transition) when
the global mean 21-cm intensity was zero (relative to the CMB). At
this time, all the other fluctuation sources are suppressed (e.g., to
linear order they are nullified, since each fluctuation gets
multiplied by the mean 21-cm intensity). This suppression includes the
velocity gradient effect, so the power spectrum is nearly isotropic
\cite{Rich} (i.e., $P_{\mu^2}$ and $r_\mu$ are both close to zero).

The analyses of 21-cm fluctuations due to fluctuations in
Lyman-$\alpha$ or X-ray radiation are quite similar
\cite{Barkana:2005b,Pritchard:2007}. When fluctuations are dominated
by one of them, in linear theory the predicted power spectrum can be
written as 
\begin{equation}
  P({\bf k},z) \approx P_{\mu^0}(k,z) = \left[ \bar{T}_b(z) 
    W_k^{\rm lin}(k,z)
  \right]^2 P_\delta(k,z) \ ,
\end{equation} 
where $\bar{T}_b$ is the mean 21-cm brightness temperature, $P_\delta$
as before is the density power spectrum, and $W_k^{\rm lin}$ is a
window function that expresses in Fourier space a convolution of
density with a radial distribution function $W_r$ in real space. They
are related by a Fourier transform relation:
\begin{equation}
  W_k^{\rm lin}(k,z) = \int_0^\infty W_r(r,z) \frac{\sin(kr)} {kr} dr\ ,
\label{Eq:Wklin}
\end{equation} 
where we consider a (randomly chosen) fixed point $P$ at a fixed $z$,
$r$ is the (comoving) distance from $P$, and $\int_0^r W_r(r',z) dr'$
equals the fraction of the total X-ray intensity at $P$ that comes
from sources up to a distance $r$ away.

Thus, in this scenario, the power spectrum can be used to measure the
radial distribution of X-ray flux. Before we test this proposition, we
note that: 1) We have dropped some overall ($k$-independent) factors,
since we only test the shape and not the normalization of the power
spectrum; 2) Even within the linear theory, we have neglected small
additional complicating factors; and 3) Non-linearity may complicate
things further.

Within our simulated cosmic volumes we measure the radial distribution
of X-ray flux, $W_r$, seen on average (i.e., averaged over many
central pixels at each redshift) \cite{FN7}. How this quantity varies with radial distance
depends on whether the radiative sources emit hard X-rays (in which
case most of the photons travel far from the sources) or soft X-rays
(in which case most of the photons are absorbed close to the sources).
We compare its transform $W_k^{\rm lin}$ from eq.~(\ref{Eq:Wklin}) to
the following quantity measured from our simulated results as in mock
21-cm observations:
\begin{equation}
  W_k(k,z) = \sqrt{\frac{\langle P({\bf k},z) \rangle}{P_\delta \bar{T}_b^2}}\ ,
\label{Eq:wk}
\end{equation}
i.e., this is the ratio of the measured isotropically-averaged 21-cm
power spectrum to the theoretically calculated density power spectrum
$P_\delta$ (which is well determined based on other cosmological
probes such as the CMB). We have included the factors of $\bar{T}_b$
(which is not directly observable) simply to make $W_k$ dimensionless,
but as mentioned, we are interested only in the shape of $W_k$ as a
function of $k$, and not in its overall normalization.

\begin{figure*}
\includegraphics[width=6.0in]{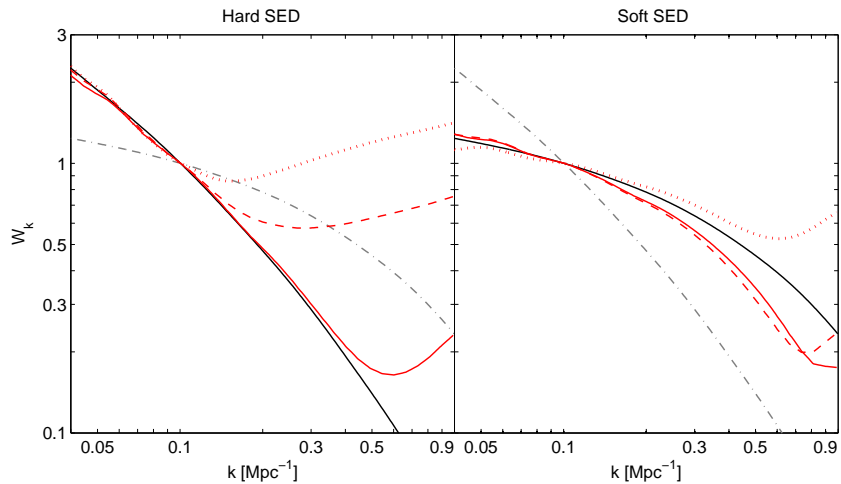}
\caption{A model-independent method for determining the spectrum of
  the X-ray sources that heated the universe. We compare $W_k$ as
  measured from mock observations to the predicted $W_k^{\rm lin}$
  from the radial distribution of X-ray flux as measured in our
  simulations. We compare our two cases of a hard X-ray SED (left
  panel) or a soft SED (right panel). We show $W_k$ at a range of
  redshifts, $z=12$ (solid red), 13 (dashed red), and 14 (dotted red)
  for the hard SED, and $z=14$ (solid red), 16 (dashed red), and 18
  (dotted red) for the soft SED. Each case is compared to $W_k^{\rm
    lin}$ (which changes little with $z$) at the central redshift
  (black curve); for contrast we also show $W_k^{\rm lin}$ from the
  other panel (grey, dot-dashed line). We only wish to compare shapes,
  so all the curves are normalized to unity at $k = 0.1$~Mpc$^{-1}$
  (but to illustrate the overall amplitude at these redshifts, we note
  that the highest fluctuation at $k=0.2$ is 3.3~mK for the hard SED
  at $z=14$, and 12.4~mK for the soft SED at $z=16$). We note that the
  slopes of $W_k$ and $W_k^{\rm lin}$ agree only on scales on which
  the 21-cm fluctuations are dominated by heating fluctuations; also
  note that in this Figure we are able to go to much larger scales
  than in Figure~\ref{Fig:P}, since the isotropically-averaged power
  spectrum can be determined more accurately in a given simulation box
  than parameters that depend on the angular variation.}
\label{Fig:Wk}
\end{figure*}

Applying this method to heating fluctuations, we find that it is
indeed possible to reconstruct the radial distribution of the X-ray
radiative background from 21-cm data (Figure~\ref{Fig:Wk}). The slope
of $W_k$ matches that of $W_k^{\rm lin}$ remarkably well near the
heating transition, i.e., at $z\sim 12-14$ for the hard SED and $z\sim
14-18$ for the soft SED. This works only on large scales (up to
$k=0.1-0.5$~Mpc$^{-1}$ depending on $z$ and the SED), where the
heating fluctuations dominate. As expected, in the case of the soft
X-rays the slope is flatter, i.e., the contribution from short
distances (high $k$) is larger. The difference between the two SEDs is
large and easily seen. Thus, measuring the slope of $W_k$ near the
heating transition reveals the hardness of the X-ray SED.

{\it \bf Conclusions.}\ \ In this Letter we have quantified for the
first time the anisotropy in the 21-cm power spectrum over the wide
range of cosmic epochs from cosmic dawn to the end of the epoch of
reionization ($z = 30-7$) using mock 21-cm observations. We show that
the anisotropy is large and it can be used as a cosmic ``clock'' to
determine the timing of cosmic events. We also show that, in the case
of X-ray heating, it is possible to reconstruct the shape of the
radial flux distribution from measurements of the 21-cm power
spectrum. This provides a new tool to constrain the nature of the
first sources of heat in the Universe.

Our findings are especially timely due to the great current interest
in observations of the 21-cm signal, including the recent first
constraints on cosmic heating from the PAPER experiment \cite{PAPER}.
Our results will be useful for the interpretation of future data from
radio telescopes such as HERA \cite{HERA} and the Square Kilometre
Array \cite{Mellema:2013}, and should make the anisotropy of the 21-cm
power spectrum a topic of major theoretical and observational
interest.

{\it Acknowledgments:}\ \ A.F.\ was supported by the LabEx ENS-ICFP:
ANR-10-LABX-0010/ANR-10-IDEX- 0001-02 PSL and NSF grant AST-1312034.
R.B.\ and A.C.\ acknowledge Israel Science Foundation grant 823/09 and
the Ministry of Science and Technology, Israel. This work was mostly
carried out during a visit by R.B.\ to UPMC Univ Paris 06.


\begin{thebibliography}{}
\bibitem{FN1} Note that here we have defined the coefficients according to their angle-averaged size
(e.g., $P_{\mu^4}$ is defined accounting for 
$\left< \mu_k^4 \right> = 1/5 $).
\bibitem{FN2} While the cited investigations suggest that the $P_{\mu^4}$
  term {\em can}\/ be reconstructed early in reionization, they did
  not include the other early fluctuation sources that we do.
\bibitem{FN3} We consider the
  previous two attempts to be much less realistic than ours. One
  \cite{Zawada14} used a full simulation technique, but could only
  resolve Milky-Way mass halos, orders of magnitude above the expected
  masses of typical galactic halos at high redshift. Another recent
  paper \cite{Ghara} reached the erroneous conclusion of a negligible
  anisotropy at early times, due to an incorrect calculation of X-ray
  heating that removed its correlation with density even though
  density fluctuations are the driver of X-ray heating fluctuations.
\bibitem{FN4} For clarity we note that when we compute
  $P_{\mu^2}^{\rm lin}$ we do not linearize any fluctuations, but
  simply assume linear separation and calculate the cross-correlation
  of density with the actual (non-linear) isotropic 21-cm
  fluctuation.
\bibitem{FN5} These changes of sign are
  fairly simple to explain \cite{Rich}. For example, ionizing a given
  region would reduce its neutral hydrogen content, reducing its 21-cm
  signal, while increasing its density would increase its 21-cm
  signal; this gives rise to a negative correlation in this case.
\bibitem{FN6} The averages in eq. (2) refer
  to averages over angles. The power spectrum itself is determined by
  averaging over a large volume or equivalently, in our calculations,
  over many simulations.
\bibitem{FN7} Note that we account
  properly for the fact that the flux is redshifted and the sources
  are time-retarded.
\end{thebibliography}
{}
\end{document}